\documentclass[12pt]{iopart}
\RequirePackage{lineno}
\usepackage{iopams}
\usepackage{graphicx}
\usepackage{wasysym}
\usepackage{threeparttable}
\begin{document}
\setpagewiselinenumbers
\modulolinenumbers[1]
\linenumbers
\title[Quantitative analyses of empirical fitness landscapes]{Quantitative analyses of empirical fitness landscapes}

\author{Ivan G Szendro$^1$, Martijn F Schenk$^{2,3}$, Jasper Franke$^1$, Joachim Krug$^{1,4}$ and J Arjan GM de Visser$^3$}

\address{$^1$ Institute for Theoretical Physics, University of Cologne, K\"oln, Germany}
\address{$^2$ Institute for Genetics, University of Cologne, K\"oln, Germany}
\address{$^3$ Laboratory of Genetics, Wageningen University,
  Wageningen, The Netherlands}
\address{$^4$ Systems Biology of Ageing Cologne (Sybacol), University
  of Cologne, K\"oln, Germany} 
\ead{krug@thp.Uni-Koeln.de}
\begin{abstract}
The concept of a fitness landscape is a powerful metaphor that offers
insight into various aspects of evolutionary processes and guidance
for the study of evolution. Until recently, empirical evidence on the
ruggedness of these landscapes was lacking, but since it  became
feasible to construct all possible genotypes containing combinations
of a limited set of mutations, the number of studies has grown to a
point where a classification of landscapes becomes possible. The aim
of this review is to identify measures of epistasis that allow a
meaningful comparison of fitness landscapes and then apply them to the
empirical landscapes to discern factors that affect ruggedness. The
various measures of epistasis that have been proposed in the literature appear to be equivalent. Our comparison shows that the ruggedness of the empirical landscape is affected by whether the included mutations are beneficial or deleterious and by whether intra- or intergenic epistasis is involved. Finally, the empirical landscapes are compared to landscapes generated with the Rough Mt.\ Fuji model. Despite the simplicity of this model, it captures the features of the experimental landscapes remarkably well.
\end{abstract}

\maketitle

\section{Introduction}
\label{intro}
\begin{figure}
\centerline{\includegraphics[width=.8\textwidth]{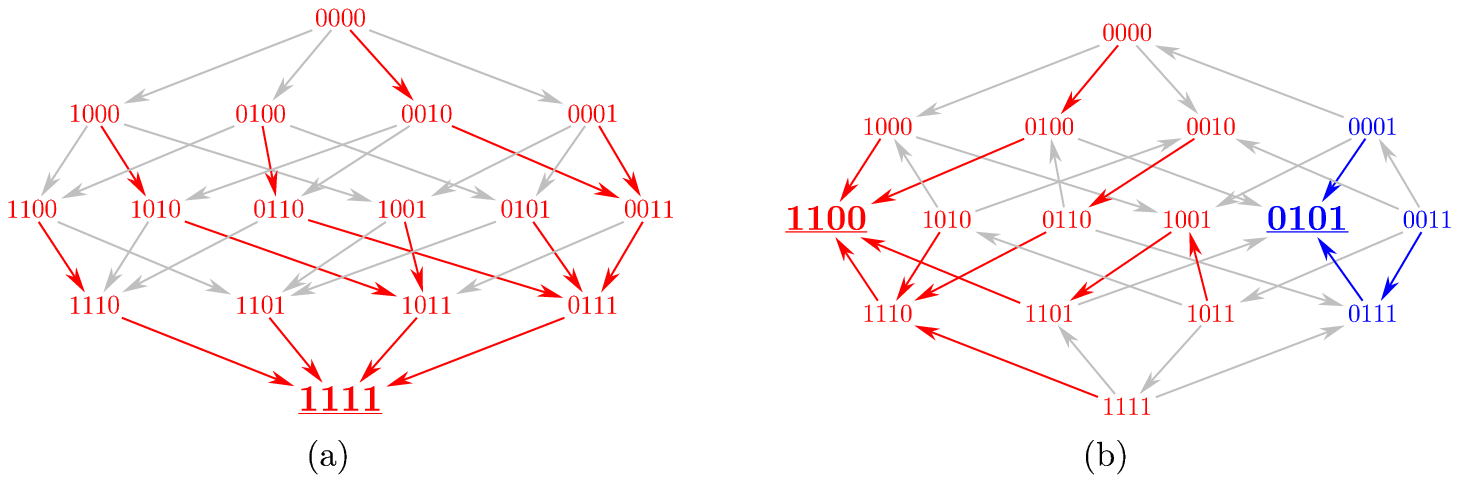}}
\caption{The figure shows two examples of empirical fitness landscapes containing all combinations of 
mutations at $L=4$ loci. Genotypes are represented  by binary sequences, where 0 (1) indicates the absence 
(presence) of the corresponding mutation. Arrows point in the direction of increasing fitness, and underlined nodes 
correspond to fitness maxima. Colored arrows point towards the fittest neighbor, forming the basins of attraction of a steepest ascent (``greedy'') adaptive walk. (a) Fitness landscape based on four beneficial mutations in the 
bacterium \textit{Methylbacterium extorquens} \cite{Chou2011} (landscape A in Tables \ref{table1}
and \ref{table2}). 
(b) Fitness landscape based on four mutations in a malaria drug resistance gene \cite{Lozovsky2009}. The fourfold mutant $\{1111\}$ confers maximal drug resistance but does not optimize the growth rate in the absence of the drug,
which is the quantity used here as a proxy for fitness (landscape D in Tables \ref{table1} and \ref{table2}).}
\label{fig:1}
\end{figure}

How genotypes map onto phenotypes is one of the central questions in biology. Developmental and systems biologists seek to understand the physical, biochemical and physiological basis of the genotype-phenotype map, while evolutionary biologists study its evolutionary causes and consequences \cite{Costanzo2010,Visser2011,Wagner2011}. To predict the evolutionary fate of a genotype it is essential to understand how genotypes map onto fitness -- the basic predictor of an organism's evolutionary success. This has led to the notion of a fitness landscape \cite{Haldane1931,Wright1932}, which is a mapping from the multidimensional genotype space to a real-valued measure of fitness. Graphical renderings often depict the fitness landscape as a surface above a two-dimensional base plane symbolizing 
the genotype space, but it is clear that such a low-dimensional representation is generally inadequate to 
provide more than a rather superficial, metaphoric description of the evolutionary process
(for an alternative visualization see fig.\ref{fig:1}). The limitations of the 
two-dimensional representation have spawned much fundamental criticism of the fitness landscape concept.
Here, rather than abandoning the concept altogether, we take the view that \textit{''fitness landscapes...should be studied in less picturesque but more quantitative ways''} \cite{Carneiro2009}. 

Within the fitness landscape metaphor, adaptation is imagined as a
hill-climbing process leading the population to a fitness peak, with
distinct roles for both natural selection and genetic drift
\cite{Jain2007,Wagner2008}. The structure of the fitness landscape can
range from smooth with few accessible peaks to rugged with multiple
peaks separated by valleys of low fitness. Whether the landscape is
smooth or rugged has important consequences for evolution
\cite{Phillips2008,Wolf2000}. For instance, the topography of the
fitness landscape affects speciation via reproductive isolation
\cite{Gavrilets2004,Kondrashov2001}, the evolutionary benefits of sex
and recombination \cite{Visser2009,Otto2009}, the evolution of genetic
robustness and evolvability \cite{Lenski2006,Nimwegen1999,Wagner2005},
and the predictability of evolution \cite{Jain2007,Lobkovsky2011,Salverda2011,Colgrave2005}.

Little is known about the factors that determine the topography of a fitness landscape, beyond the general notion that \textit{epistasis} is involved. The term epistasis as defined by Fisher \cite{Fisher1918} includes all deviations from the additive effects of alleles at different loci, and is usually considered for two alleles only. To understand the role of epistasis in shaping the structure of fitness landscapes we need to distinguish between \textit{magnitude} and \textit{sign} epistasis \cite{Weinreich2005}. Magnitude epistasis is present when the fitness effect of a mutation at a given locus has a definite sign (beneficial or deleterious) irrespective of the alleles at other loci, while
the magnitude depends on the genetic background. Magnitude epistasis
does not constrain accessibility of mutational trajectories in an
absolute sense and only
affects the curvature of a landscape, which can be quantified  by a
quadratic regression of mean fitness on mutation number. On the other
hand, sign epistasis occurs when mutations are beneficial in some
genetic backgrounds, but not in others. Hence, the sign (positive or
negative) of an allelic effect changes with the presence of an allele
at another locus. Sign epistasis causes pathways to become
inaccessible by natural selection and thus introduces ruggedness into
the landscapes \cite{Wagner2005,Weinreich2005}. A special case of sign
epistasis, called reciprocal sign epistasis, occurs when the sign of
both alleles' fitness effects changes with a change of alleles at the
other locus. Reciprocal sign epistasis is a prerequisite for the occurrence of multiple fitness peaks \cite{Poelwijk2007,Poelwijk2011}. 

A related but distinct classification of epistatic interactions
discerns between unidimensional and multidimensional epistasis
\cite{Visser2011,Kondrashov2001}. In the unidimensional case fitness can be
written in terms of a scalar function of the genotype, such as 
the number of loci carrying a mutation, whereas multidimensional
epistasis includes all allelic interactions. 
Generic fitness landscapes are expected to display
both sign epistasis and multidimensional epistasis. Nevertheless,
in empirical studies a unidimensional fitness function has often been \textit{assumed} by averaging over
different allelic combinations carrying the same number of mutations. 
Such an approach can be misleading, because a seemingly additive
unidimensional fitness landscape may result from the cancellation of 
multidimensional epistatic interactions \cite{Visser1997}.

In theory, with full knowledge of a fitness landscape, one overlooks
all possible evolutionary pathways connecting two genotypes, and would
be able to determine the likelihood that particular pathways are
taken. This would render evolution predictable in a restricted (i.e.,
\textit{a posteriori}) sense. However, these predictions are valid for
a specific combination of genotype and environment, and also depend on population dynamic parameters such as population size and mutation rate. Another limitation is that one can only study a tiny part of sequence space explicitly, because the number of genotypes grows beyond comprehension with the number of loci considered. Even for a single gene of 1000 base-pairs and when allowing only point mutations, the number of possible genotypes ($4^{1,000}$) is larger than the total number of particles in the universe, as Sewall Wright \cite{Wright1932} realized. This scale problem has two immediate implications. First, it emphasizes the fundamentally stochastic nature of evolution, given how little of genotype space has been probed by life since it exists. Second, if we are to use the growing amount of information about the genetic make-up of organisms to understand and ultimately predict evolution, we need to invoke models of fitness landscapes parametrized by empirical observations.

The purpose of this review is to compare the topographies of empirical fitness landscapes that recently have been published. Before doing so, we briefly survey the main models of fitness landscapes in which ruggedness can be tuned, as well as the different approaches to study fitness landscapes empirically. Recent efforts have been directed towards constructing all $2^L$ possible genotypes containing combinations of a limited set of $L$ mutations, and measuring their fitness or a proxy thereof \cite{Visser2011}, see fig.\ref{fig:1}
for two examples. First, we compare different measures of ruggedness and sign epistasis derived from the available landscapes, and find that these correlate well and thus appear to be equivalent. Second, using these robust measures of ruggedness we can compare the ruggedness of different empirical landscapes despite of methodological differences and the variety of biological systems involved. We find that those landscapes built from mutations that are known to have a combined beneficial effect are less rugged than those built from mutations that are selected without regard to their combined effect, in particular when they are deleterious. Third, we compare the empirical fitness landscapes to a simple statistical one-parameter model, the Rough Mt. Fuji model, which combines a linear fitness trend with uncorrelated random fitness variations. We find that this model captures the features of the empirical landscapes surprisingly well.

\section{Fitness landscape models}
\label{sec:2}
Until recently empirical information about the structure of fitness landscapes was largely unavailable, 
and the number of studies is currently still small. Therefore, past studies of fitness landscapes have been mostly restricted to theoretical work. The models proposed in this context are based on very different - and sometimes even contradicting - intuitions. In this section, we give a brief overview of the most popular models.
Throughout we represent genotypes by binary strings $\vec{\sigma} = (\sigma_1, \sigma_2, ..., \sigma_L)$ of length
$L$, where $\sigma_i = 0$ (1) if the mutation at the $i$-th locus is absent (present), compare to fig.\ref{fig:1}.

\subsection{Kauffman's $LK$ model}
If a given gene $A$, when expressed, produces an essential protein that requires the presence of another protein produced by gene $B$ to function properly, these two genes interact epistatically. When gene $A$ is expressed independently of gene $B$, an organism with a defective mutation in gene $B$ incurs the cost of producing the protein from gene $A$ without experiencing its beneficial effect. The mutation in gene $B$ can thus change the fitness contribution of gene $A$ from beneficial to deleterious, resulting in sign epistasis. The main motivation of the $LK$ model\footnote{The model 
was originally introduced under the name `$NK$ model'. The present
designation, first adopted in \cite{Franke2011}, is motivated by the
fact that the letter $N$ denotes population size in much of the
population genetic literature. The number of loci is therefore
more appropriately named $L$.} as proposed by Kauffman and Weinberger \cite{Kauffman1989,Kauffman1993} is to capture such strong, sign-epistatic effects of single mutations in interacting genes on fitness in a statistical sense, without attempting a detailed biochemical description. 

The interactions are modeled as evenly spread across the entire
genome. The genome is represented as a binary sequence of length $L$
and the interactions take the form of a set of sites called
interaction partners $\vec{\nu}_i\equiv\{i,\nu_{i,1},\nu_{i,2}, \dots,
\nu_{i,K} \}$ associated with each site $i$ of the genome. The number
$K$ of interaction partners is kept constant. How partner sites are
assigned has implications for search strategies for the global optimum
on the landscapes \cite{Weinberger1996}, but most often the interaction partners are chosen by picking them uniformly and independently at random (making sure that no site appears twice in a given set $\vec{\nu_i}$).  
The fitness of an organism with genome $\vec{\sigma}$ is then the sum of the individual fitness contributions, 
\begin{equation}\label{lk_fitness_sum}
  f(\vec{\sigma})=\sum_{i=1}^{L}f_i(\{\sigma_j\}_{j\in\nu_i}). 
\end{equation}
The single site contributions $f_i$ are independent and identically distributed random variables
(i.i.d.\ RV's) associated with each of the $2^{K+1}$ possible states of the argument. If a mutation hits part of the sequence that does not appear in the argument of $f_{i}$, i.e. if the mutation does not involve the site $i$ or any of the associated partner sites in $\vec{\nu}_i$, the fitness contribution $f_i$ remains unchanged, otherwise it is replaced by an independent random number.

For $K=0$, each contribution $f_i$ can only take two possible values, corresponding to $\sigma_i=0$ or $1$, respectively. Thus each site has one state that is more beneficial than the other\footnote{Provided the fitness values are drawn from a continuous distribution, the probability of a tie is zero.} and since the fitness of a sequence is the sum over the site contributions, the global optimum is at the state with all sites in their `beneficial' position. The global optimum can be reached from any initial configuration by mutating sites into their beneficial 
state in a random order, which implies that all mutational pathways are accessible and sign epistasis is absent. 
Conversely, when $K=L-1$, the entire sequence appears in the argument of each single site contribution. Thus any mutation replaces the sum (\ref{lk_fitness_sum}) by a sum over a different set of i.i.d.\ RV's, which is equivalent to replacing one i.i.d.\ RV by another. The number of sites in the partner set, denoted by $K$, thus allows one to tune the 
strength of epistatic interactions from the non-epistatic limit $K=0$ to the maximally epistatic case $K=L-1$. 

\subsection{Rough Mount Fuji Models}
\label{sec:RMF}
The $K=0$ limit of the $LK$ model can be compared to a smooth (though not necessarily symmetric) mountain much like the Mt. Fuji volcano in Japan. This type of fitness landscape is therefore sometimes referred to as `Mt Fuji' landscape. The other extreme ($K=L-1$) corresponds to a maximally rugged landscape of independent fitness contributions and is referred to as `House of Cards' (HoC) landscape \cite{Kingman1978}\footnote{Interpreting the genotype sequences as spin 
configurations, the HoC landscape becomes equivalent to Derrida's Random Energy Model of spin glasses, and 
the $LK$-model is a close relative of the $p$-spin model \cite{Derrida1981}.}. Intermediate values of $K$ correspond to an intermediate degree of ruggedness. An alternative way to obtain landscapes with intermediate ruggedness is to pick one genotype $\vec{\sigma_0}$ as point of reference and then impose an external `fitness field' of strength $c$ favoring this reference configuration on top of random i.i.d.\ contributions. Then the fitness of a genotype $\vec{\sigma}$ is given by 
\begin{equation}\label{rmf_fitness_sum}
f(\vec{\sigma})=\eta(\vec{\sigma})-cd(\vec{\sigma_0}, \vec{\sigma}),
\end{equation}
where $d(\vec{\sigma_0}, \vec{\sigma})$ denotes the Hamming distance between the two configurations and 
$\eta(\vec{\sigma})$ is a random variable picked independently for each genotype. This is a simplified version of the Rough Mount Fuji model as originally introduced in \cite{Aita2000}, which has also been used in \cite{Franke2011} (see also \cite{Lobkovsky2011}). If the i.i.d.\ part of the fitness fluctuates on a scale $a$, the landscape will be dominated by the external field and appear like a smooth landscape for $c/a\gg 1$, while the random contributions will dominate for $c/a\ll 1$, making the landscape appear maximally rugged. Note that the model assumes that the mean fitness profile (averaged over
the random fitness component $\eta(\vec \sigma)$) is linear, and thereby ignores unidimensional magnitude epistasis. However, since our main interest is in measures of landscape ruggedness, the mean curvature of the landscape is not relevant. In section \ref{subsec:3.2}, we will compare measures of epistasis and landscape ruggedness for empirical data to those obtained for landscapes constructed with the RMF model (\ref{rmf_fitness_sum}), choosing a Gaussian distribution with standard deviation $a$ for the i.i.d.\ random variables $\eta(\vec{\sigma})$. 

\subsection{Neutral Models}
A different intuition than used for the $LK$ and RMF models is that
the actual fitness matters little compared to the question whether or
not a given organism is viable at all. The genome is composed of a
large number of mutually interacting elements and a random mutation in
any given gene can alter gene function up to the point where a gene does no longer function. It has therefore been postulated that fitness landscapes are dominated by large valleys of lethality and extended ridges of viability \cite{Gavrilets2004}. 
In the simplest setting each genotype $\vec{\sigma}$ has either fitness $1$  (i.e.\ is viable) with probability $p$ or has fitness $0$ (not viable) with probability $1-p$, independent of other states. The resulting fitness landscape is then equivalent to a realization of the site percolation problem \cite{Stauffer1991} on an $L$-dimensional hypercube \cite{Gavrilets1997}. This type of model can be combined with the models described in the preceding subsections by
introducing a fraction of non-viable genotypes in addition to the epistatically interacting viable genotypes, 
see \cite{Franke2011}. 

\subsection{Models with explicit phenotypes}
The models described so far intend to incorporate known aspects of the biochemical or biological interactions shaping the fitness of a given organism while keeping the number of parameters to a minimum. Another strategy is to explicitly incorporate physical, chemical and biological mechanisms underlying epistasis into an explicit genotype-phenotype map \cite{Khatri2009}. Such models have been based, for example, on the thermodynamics of RNA secondary structure \cite{Fontana1993} or the biophysics of binding between a transcription factor and its binding site \cite{Mustonen2008}.  While the development of such models constitutes an active branch of research, they are too complex and specific for the type of analyses that are of interest in the context of this review.

\section{Empirical studies of fitness landscapes}
\label{sec:3}
Several approaches have been used to infer topographical properties of real fitness landscapes from empirical observations; these can be roughly classified into three  categories. Studies in the first category use the repeatability of adaptation observed in microbial evolution experiments to qualitatively assess local ruggedness of fitness landscapes. Studies in the second category focus on detecting sign epistasis between mutations to infer local ruggedness. The third category includes a limited, but growing number of studies that explicitly quantify the multidimensional fitness landscape by considering all $2^L$ combinations of a small set of $L$ mutations. The topographical information revealed by the first two approaches is necessarily limited, but reflects the contribution of a large number of mutations, while the third category yields more detailed information, but from a tiny predefined part of genotype space. In the next section, we will briefly review several studies from the first two categories, and then present a more extensive analysis of available studies from the third category.

\subsection{Empirical support for global ruggedness}
By allowing replicate populations of microbes to evolve under identical conditions in the laboratory, the dynamics and repeatability of adaptation can be quantified and used to infer the general ruggedness of the fitness landscape involved \cite{Colgrave2005,Elena2003}. One expectation is that a rugged landscape leads to a stronger and more sustained divergence of fitness trajectories than a smooth landscape. This has been found when comparing bacteria evolving in a structured and a non-structured habitat \cite{Korona1994}, or in a complex relative to a simple nutrient environment \cite{Rozen2008}. Another expectation is that only on rugged landscapes the ability to adapt depends on the local mutational neighborhood of a genotype. In contrast, all genotypes except the globally optimal genotype are able to adapt on a smooth single-peaked landscape. Support for this expectation comes from a study with RNA bacteriophage $\phi$6, where only one of two related genotypes was able to adapt under identical conditions \cite{Burch2000}, and from a study with HIV-1 where adaptation to one host-cell type could only be realized indirectly through adaptation to another host-cell environment \cite{Opijnen2007}. Another prediction for rugged landscapes is that higher levels of adaptation diminish the ability to adapt to different niches, which was found in a study with biofilm-producing bacteria \cite{Buckling2003}. Finally, the short adaptive walks found in recent experiments with fungi \cite{Gifford2011,Schoustra2009} also suggest that their fitness landscape is rugged \cite{Orr2003,Neidhart2011}.

Attempts to infer topographical information from the dynamics and repeatability of adaptation necessarily suffers from being non-systematic. Because such studies reveal only those parts of the fitness landscape that have actually been probed, they are unable to quantify the ruggedness of the landscape. For instance, the observed adaptive dynamics may suggest that there are no strong epistatic constraints, while the population may have traveled along a rare ridge of high fitness within a rugged landscape. Conclusions also depend on the type of mutations used by evolution, which are specific for the population dynamic regime that prevailed. For instance, in large populations where clonal interference plays a major role, large-effect mutations will dominate \cite{Park2010} and their epistatic properties may be different from smaller-effect beneficial mutations, or even neutral or deleterious mutations that may contribute under different conditions \cite{Visser2011}. On the other hand, these approaches may probe a more extended area of genotype space than the more systematic approach of mutant construction involving a predefined and small set of mutations, which is discussed in sect.\ \ref{subsec:3.2}. 

Epistasis has a clear link to ruggedness of fitness landscapes. Several studies confirmed the role of epistasis in causing adaptive
constraints and local ruggedness by using isolated or constructed
mutants in replay experiments to test their evolutionary consequences
\cite{Blount2008,Salverda2011,Woods2011}. Studies which examine pairwise
interactions within sets of mutations often detect sign
epistasis, and also provide information regarding its frequency
\cite{Visser2011}, implying that local ruggedness is not uncommon. For
example, a study on beneficial mutations that increase the growth rate
of the ssDNA microvirid bacteriophage ID11 found significant evidence
for sign epistasis in six out of 18 constructed combinations
\cite{Rokyta2011}. A study by Sanjuan et al. \cite{Sanjuan2004} on
vesicular stomatitis virus identified five out of 15 cases in which
the combination of two mutations was less fit than either of the
single mutants. Apart from studies that focus on a relatively small
number of well-characterized mutations, the ubiquity of (pairwise) epistatic
interactions has also been documented in recent genome-wide
surveys \cite{Costanzo2010,Hinkley2011}. Mutation combinations for
which sign epistasis is identified point to the local ruggedness of
the fitness landscape, but do not reveal the global structure of the
landscape \cite{Kondrashov2001}.

\subsection{Explicit low-dimensional fitness landscapes}
\label{subsec:3.2}

\begin{table} 
\caption{General characteristics of the empirical fitness landscapes included in this review. The table lists the number of loci involved, the number of available genotypes, the fitness (proxy) that is measured for each combination, and the type of mutations included in the landscapes. The organism is indicated (in \textit{italics}) when landscapes are based on genome-wide mutations or the gene name is provided (in upright letters) when mutations are located on a single gene. Columns 6 and 7 indicate whether the included mutations were known or expected to be beneficial or 
deleterious, individually and/or in combination. The results of our quantitative analyses for the landscapes A--J are shown in table \ref{table2}.}
\label{table1}       
\centering
\begin{threeparttable}
\centering
\begin{tabular}{lllllllll}
\hline\noalign{\smallskip}
 ID & System   & $L$ & Available & Fitness & Direction of& Known & Ref.\\
&   (\textit{organism}/gene) &  & combinations & (proxy) & mutations& effects&\\
\noalign{\smallskip}\hline\noalign{\smallskip}
A &  \textit{Methylobacterium} & 4 & 16/16 & Growth rate & Beneficial & Combined & \cite{Chou2011} \\
 &   \textit{extorquens} &  &  &  &  &  &  \\
B &  \textit{Escherichia} & 5 & 32/32 & Fitness & Beneficial & Combined & \cite{Khan2011} \\
 &   \textit{coli} &  &  &  &  &  &  \\
C-D &  Dihydrofolate & 4 & 16/16 & Resistance/ & Beneficial\tnote{a} & Individual/ & \cite{Lozovsky2009} \\
 &   reductase &  &  & Growth rate &  & Combined &  \\
E &  $\beta$-lactamase & 5 & 32/32 & Resistance & Beneficial & Combined\tnote{b} & \cite{Weinreich2006} \\
F &  $\beta$-lactamase & 5 & 32/32 & Resistance & Beneficial\tnote{c} & Combined\tnote{c} & \cite{Tan2011} \\
G &  \textit{Saccharomyces} & 6 & 64/64 & Growth rate & Deleterious & Individual & \cite{Hall2010} \\
 &   \textit{cerevisiae} &  &  &  &  &  &  \\
H &  \textit{Aspergillus} & 8 & 186/256\tnote{d} & Growth rate & Deleterious & Individual & \cite{Franke2011} \\
 &   \textit{niger} &  &  &  &  &  &  \\
I-J &  Terpene synthase & 9 & 418/512\tnote{d} & Enzymatic & -- & -- & \cite{OMaille2008} \\
 &   \textit{} &  &  & specificity\tnote{e}  &  &  &  \\
-- &  Dihydrofolate & 5\tnote{f} & 29/48\tnote{d} & Resistance/ & Beneficial & Individual/ & \cite{Brown2010} \\
 &   reductase  &  &  & Growth rate &  & Combined &  \\
-- &  Dihydrofolate & 5\tnote{f} & 29/48\tnote{d} & Resistance/ & Beneficial & Individual/ & \cite{Costanzo2011} \\
 &   reductase  &  &  & Growth rate &  & Combined &  \\
-- &  HIV-1 envelope & 7 & 56/128\tnote{g} & Infectivity & Beneficial & Individual/ & \cite{DaSilva2010} \\
 &  glycoprotein gp120     &  &  &  &  & Combined &  \\
-- &  Isopropylmalate & 6\tnote{f} & 164/512\tnote{g} & Performance/ & -- & -- & \cite{Lunzer2005} \\
 &    dehydrogenase &  &  & Fitness &  &  &  \\
\noalign{\smallskip}\hline
\end{tabular}
\begin{tablenotes}
\footnotesize
\item [a] The mutants were chosen to maximize drug resistance but do not optimize the growth rate in the absence of the drug.
\item [b] The highly resistant genotype resulted from gene-shuffling, which implies that an accessible pathway between the wildtype and this mutant did not necessarily exist.
\item [c] The same mutants as in \cite{Weinreich2006} were studied
  with respect to piperacillin+inhibitor resistance. Due to the strong
  negative correlation between cefotaxime- and piperacillin+inhibitor-resistance the wildtype was expected to be exceptionally fit.  
\item [d] The remaining combinations were missing, either by chance or because the corresponding phenotypes are not viable. 
The studies \cite{Brown2010,Costanzo2011} were excluded from further analysis because of the large number of missing combinations.
\item [e] The study considers mutational pathways connecting two terpene synthases, TEAS and HPS. Enzymatic specificity is the relative 
proportion of the natural product of TEAS (landscape I) and HPS (landscape J) among the total catalytic output of the mutated enzymes.   
\item [f] More than one mutation was included for at least one locus, hence the number of possible combinations
is larger than $2^L$. 
\item [g] The remaining combinations were not engineered. These studies were excluded from further analysis because of  the large number of missing combinations.

\end{tablenotes}
\end{threeparttable}
\end{table} 

The existence of sign epistatic interactions between mutations reveals
that landscape topography can be rugged, but a more systematic
approach is required to quantify the degree of ruggedness of fitness
landscapes, and to determine how this constrains evolution. Given the large number
of publications on fitness landscapes, the number of studies on
empirical fitness landscapes is remarkably small. Full information is
available when the fitness of all $2^L$ combinations of a set of $L$
mutations is known. At present, available empirical landscapes stem
from a variety of systems and involve small numbers (i.e., 4-9) of
mutations (see table \ref{table1}). In reality, adaptation proceeds by
selection on all possible mutations in the genome and is not
necessarily limited to such a small subset. These landscapes thus only
offer a glimpse of the ruggedness within the immense genotype
space. Given this limitation, we cannot compare ruggedness between
different empirical fitness landscapes without a clear view on which
mutations are involved and which part of genotype space is being
mapped. As it turns out - despite of their low number - the available
landscapes are rather different in several aspects, and include
mutations in single genes versus whole genomes, with fitness effects 
that in some cases are known \textit{a priori} to be beneficial or deleterious
(for individual mutations or for the combination studied) and in other
cases emerge only \textit{a posteriori}. For each of the empirical
landscape studies included in our analyses (and a few which
we did not include), table \ref{table1} summarizes the main characteristics.

\subsubsection{Quantitative  measures of landscape ruggedness and epistasis.}
\label{subsec:3.2.2}
Various statistical measures have been proposed to quantify the
ruggedness of fitness landscapes. Most studies focused on
different aspects of landscape topography and a variety of measures has consequently
been applied. Here, we aim to analyze all landscapes using a common
and standardized selection of measures. This enables a comparison
between landscapes, but also allows us to verify whether model
landscapes actually capture the topography of the empirical
landscapes. Furthermore, we explore the correlations between the
different measures of ruggedness to see whether they are equivalent or
yield complementary information about the topography of real fitness
landscapes. In total, we use six measures for our analyses:

\begin{itemize}
\item[(1)] The \textit{roughness to slope ratio}, $r/s$, was
  introduced in \cite{Aita2001} and used in
  \cite{Carneiro2009,Lobkovsky2011}. This ratio measures how well the
  landscape can be described by a linear model, which corresponds to
  the purely additive (non-epistatic) limit. It is obtained by fitting
  a multidimensional linear model to the empirical fitness landscape
  by means of a least-square fit. The linear model is
\begin{equation}
 f^{\mathrm{fit}}(\vec{\sigma})=a^{(0)}+\sum_{j=1}^{L}a^{(1)}_j
   \sigma_{j}, \label{linfit}
\end{equation}
where the parameters $a^{(0)}$ and the $a_j^{(1)}$'s are fitted. The mean slope is
\begin{equation}
\label{slope}
s=\frac{1}{L} \sum_{j=1}^L |a_j^{(1)}| 
\end{equation}
and the roughness is defined by   
\begin{equation}
\label{roughness}
r=\sqrt{2^{-L}\sum_{\vec{\sigma}} (f(\vec{\sigma})-f^{\mathrm{fit}}(\vec{\sigma}))^2}.
\end{equation}
The higher $r/s$, the higher the deviation from the linear model and
the more epistasis is present in the landscape. For example, a purely
additive landscape has $r/s=0$, while for the HoC model $r/s \to \infty$
for $L \to \infty$.

\item[(2)] A versatile set of measures is provided by the
  \textit{Fourier analysis} of fitness landscapes introduced in
  \cite{Stadler1996}. Here, the fitness landscape is expanded in terms
  of the eigenvectors of the Laplacian on the underlying genotype
  network (in our case, the $L$-dimensional hypercube). The Laplacian is defined as $\Delta = A-L\mathbf{1}$, where
  $A$ is the adjacency matrix and $\mathbf{1}$ is the unit matrix of
  dimension $2^L\times 2^L$. Note that this matrix has $2^L$
  eigenvalues, and thus eigenstates, but that the $n$-th non-negative
  eigenvalue comes with a multiplicity given by the binomial
  coefficient ${L \choose n}$, such that the eigenvalues take only
  $L+1$ different values. 
The expansion of the fitness landscape into the eigenvectors
$\Lambda^{jn}$ of $\Delta$ is equivalent to an expansion in terms
corresponding to epistatic interactions of different orders, i.e.\
\begin{eqnarray}
f(\vec{\sigma})&=&b_{1}^{(0)}\Lambda_{\vec{\sigma}}^{10}+\sum_{j=1}^{{L \choose 1}}b_j^{(1)}\Lambda_{\vec{\sigma}}^{j1}+\sum_{j=1}^{{L \choose 2}}b_j^{(2)}\Lambda_{\vec{\sigma}}^{j2}+\ldots+b_1^{(L)}\Lambda_{\vec{\sigma}}^{1L} \nonumber\\
&=&\tilde{a}^{(0)}+\sum_{j=1}^{L}\tilde{a}_{j}^{(1)}\tilde{\sigma}_{j}+\sum_{\stackrel{j,k=1}{j > k}}^{L} \tilde{a}_{jk}^{(2)}\tilde{\sigma}_{j}\tilde{\sigma}_{k}+\ldots+\tilde{a}^{(L)}\tilde{\sigma}_{1}\tilde{\sigma}_{2}\ldots\tilde{\sigma}_{L},    
\end{eqnarray}
where the $b$'s and the $\tilde{a}$'s are the coefficients of the expansion, and  
we have introduced symmetric `spin' variables ${\tilde \sigma}_{j} = 2
\sigma_j - 1 = \pm 1$ for convenience \cite{Neher2011}. Note that the
$n$-th sum in the upper expression is equal to the $n$-th sum in the
lower one. The term $b_{1}^{(0)}\Lambda_{\vec{\sigma}}^{10}=\tilde{a}^{(0)}$ is a constant
which yields no information about epistasis. The second term sums the
contributions in the directions of the eigenvectors corresponding to
the second smallest eigenvalue and describes the additive,
non-epistatic, part of the fitness landscape. The remaining terms describe epistatic interactions of
increasing order. Defining
\begin{equation}
\label{Fn}
F_{n}=\frac{\beta_n}{\sum_{j=1}^L \beta_j} \;\; \mathrm{with} \;\;
\beta_n=\sum_{j=1}^{{L \choose n}}(b_{j}^{(n)})^2, \;\; n=1,...,L,
\end{equation}
we obtain measures for the contributions of epistatic interactions of
different order to the fitness landscape. The $F_{n}$ are
normalized to add up to 1, $\sum_{n=1}^L F_n = 1$. If one is only
interested in the total contribution of epistasis one should take the
sum over all terms corresponding to the interaction part, yielding the
epistasis measure
\begin{equation}
\label{Fsum}
F_\mathrm{sum}=\sum_{j=2}^{L} F_j. 
\end{equation}
For a purely additive landscape $F_1=1$, and thus
$F_\mathrm{sum}=0$. For a completely random (HoC) landscape
$F_1\rightarrow 0$ for $L\rightarrow\infty$. When interested in the
contributions of second, third or higher-order interactions one can
analyze the terms $F_2$, $F_3$, etc. separately. Note that the Fourier
analysis described here is equivalent to an analysis of variance
(ANOVA) commonly used by biologists, which was employed in
\cite{Visser1997} to estimate the contribution of main effects ($F_1$)
and all possible interactions ($F_2$, $F_3$, $F_4$ and $F_5$ summing
up to $F_\mathrm{sum}$) among two sets of five mutations in the fungus \textit{Aspergillus niger}.    

\item[(3)] A frequently used measure of landscape ruggedness is the
  \textit{number of local fitness maxima} $N_\mathrm{max}$, which
  exceeds unity only in the presence of reciprocal sign epistasis
  \cite{Poelwijk2011}. For the HoC model it is easy to see that
  $N_\mathrm{max} = \frac{2^L}{L+1}$ on average \cite{Kauffman1987,Kauffman1993},
  while the maximal possible value for any (binary) fitness landscape
  is $N_\mathrm{max} = 2^{L-1}$
  \cite{Haldane1931,Rosenberg2005}. Asymptotic expressions for the
  mean number of local maxima have been derived for the $LK$-model
  \cite{Weinberger1989,Durrett2003,Limic2004} as
  well as for the RMF model \cite{Neidhart2011b}.  
Note that, like all quantifiers that only depend on the ordering of
fitness values, $N_\mathrm{max}$ is insensitive to magnitude epistasis.

\item[(4)] While the quantities introduced so far are global measures
  of ruggedness, it is also of interest to characterize epistatic
  interactions locally. As a convenient \textit{local measure of
    epistasis} we examined all pairs of genotypes in the landscape
  with a Hamming distance of 2, and counted the fraction
  $f_\mathrm{s}$ of local
  motifs showing `simple' sign epistasis (i.e., the effect sign of a
  mutation at locus $i$ depends on the state of locus $j$ but not vice versa), and the fraction
  showing reciprocal sign epistasis, $f_\mathrm{r}$
\cite{Poelwijk2007}. For a HoC landscape, the expected values for
these quantities are $f_\mathrm{s}=f_\mathrm{r}=1/3$, while for a
purely additive landscape both vanish.

\item[(5)] Several recent studies apply measures of epistasis which are
  based on the notion of \textit{selectively accessible pathways},
  which are connected paths of single step mutations along which
  fitness increases monotonically
  \cite{Weinreich2005,Weinreich2006,Poelwijk2007,Franke2011,Lobkovsky2011}. 
Of particular interest are the direct (shortest) paths to the global
fitness maximum, since they provide a clear signature for the presence
of sign epistasis: in a landscape without sign epistasis, all paths from an arbitrary genotype to the global maximum
are accessible, while at least some of these paths become inaccessible in the presence of sign epistasis \cite{Weinreich2005}. Following
\cite{Weinreich2006,Franke2011} we count the \textit{number of
crossing accessible paths} $N_\mathrm{cp}$ that
lead to the fittest genotype starting from the reversal (antipodal)
genotype at distance $L$. For purely additive landscapes $N_\mathrm{cp}=L!$, while $N_\mathrm{cp}=1$ on average for a landscape compatible with
the HoC model, independent of the size $L$ of the landscape \cite{Franke2011}. 

\item[(6)] Besides the number of crossing paths $N_\mathrm{cp}$
  introduced above, there are other estimators of ruggedness
  and epistasis that rely on counting the number or length of
  accessible paths, e.g., the length and number of paths with a monotonic increase in fitness from
  the genotype with the \textit{lowest fitness} to the global optimum
  \cite{DePristo2007}. Some measures allow for detours while others do not. Measures that include
  neutral or double mutations into paths have also been applied
  \cite{Roy2009}. Other definitions do not take the location of
  the starting or endpoints into account, but ask for the length of the longest
  path that always leads from a state to its fittest neighbor (greedy walks)
  \cite{Orr2003,Rosenberg2005,Stadler2002} or that only admits states with
  exactly one fitter neighbor along the path
  \cite{Achaz2011}. While these path measures can yield interesting information about
evolutionary dynamics, they are often less suitable to quantify
epistasis because they correlate non-monotonically with 
conventional measures of landscape ruggedness. As an example, we
consider the ratio $f_\mathrm{mm}$ between the number of accessible
paths from the least fit to the fittest state of the landscape
divided by the number of such paths accessible on a purely additive
landscapes, allowing for arbitrary detours. 
In fig.\ \ref{fig:2} we plot $f_\mathrm{mm}$ obtained from simulations
of the RMF model (sect.\ \ref{sec:RMF}) for a range of values of
the slope $c$, while fixing the fluctuation parameter $a=0.1$. Recall
that the increase of $c$ from 0 to $\infty$ corresponds to the
transformation from a completely rugged to a perfectly additive
landscape. Thus, the amount of epistasis in the landscape decreases
monotonically with increasing $c$, while $f_\mathrm{mm}$ shows a
pronounced maximum at an intermediate value of $c$, i.e., the
dependence of $f_\mathrm{mm}$ on the amount of epistasis is
non-monotonic.  
We nevertheless include $f_\mathrm{mm}$ in our analyses to emphasize that epistasis
does not only imply adaptive constraints, and may sometimes even
promote evolvability by allowing detours. Such detours are not
accessible in purely additive landscapes, and may lead to
$f_\mathrm{mm} > 1$ (see also \cite{DePristo2007,Roy2009}). In contrast, the
other path-dependent quantity $N_\mathrm{cp}$ does have a monotonic
dependence on epistasis parameters like $c$ in the RMF model, or $K$ 
in the $LK$ model \cite{Franke2011}. 
\end{itemize}




\begin{figure}
\centerline{\includegraphics[width=.5\textwidth]{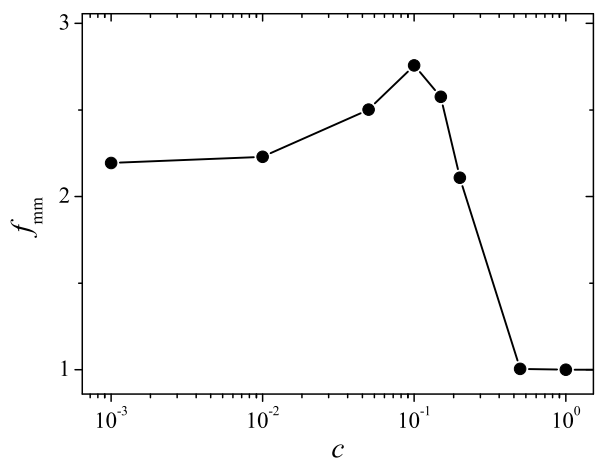}}
\caption{The number of fitness monotonic (selectively accessible) paths 
from the global fitness minimum to the global fitness maximum, divided
by the corresponding number in a 
non-epistatic landscape, $f_\mathrm{mm}$, is plotted vs.\ the slope of the RMF model landscape (\ref{rmf_fitness_sum}), $c$. Measurements were carried out on landscapes of size $L=4$ and averaged over 10,000 realizations of the landscape for each choice of $c$. Note that $f_\mathrm{mm}$ depends non-monotonically on $c$.}
\label{fig:2}       
\end{figure}

\subsubsection{Standardizing the data sets.}

Before presenting the data, one should note that the expected values of the above quantities, for a given amount of epistasis, may depend on the size of the underlying landscape. In general, we lack analytical predictions on how landscape size affects our measures, and we therefore restrict the analysis to \textit{subgraphs} of the same size. Subgraph analysis of fitness landscapes was introduced in \cite{Franke2011} as a means to probe the effect of the mutational distance scale within a fitness landscape. Here a subgraph of size $m$ is the hypercube spanned by all $2^m$ combinations of $m$ out of $L$ mutations. For landscapes of size $L > 4$  we calculated the topographic measures for all subgraphs of size $m=4$ that contained at least eight viable and known states and averaged the values over the subgraphs. For the landscapes of size $L=4$ (A, C, and D) the calculated values refer to the complete landscape. 

\begin{figure}
\centerline{\includegraphics[width=.5\textwidth]{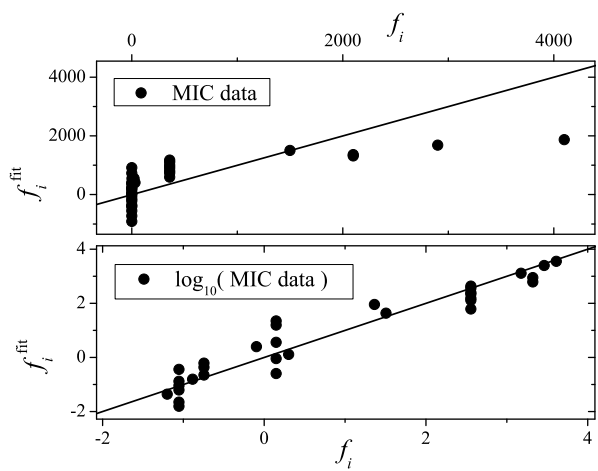}}
\caption{The measured fitness values, $f_i$,  for the
  $\beta$-lactamase resistance landscape E \cite{Weinreich2006} are plotted vs.\ the fitness values obtained from the fit with the model described by equation (\ref{linfit}), $f_i^\mathrm{fit}$. For a perfect fit, all dots would lie on the straight line with slope 1. The upper panel uses the measurements taken directly from \cite{Weinreich2006}. A systematic deviation from the straight line is observed. The lower panel uses the logarithms of the measurements; no systematic deviations are observed.}
\label{fig:3}       
\end{figure}

Furthermore, one should keep in mind that the fit to the
multi-dimensional linear model and the Fourier analysis presume that
mutations interact additively in absence of epistasis. How the effects
of mutations add up in the interaction free case will, however, depend
on the quantity one measures as a proxy of fitness. For example, when
the linear model is fitted to the landscape E of \cite{Weinreich2006},
which is based on measures of the minimal inhibitory concentration
(MIC) of an antibiotic, systematic deviations from the measured
landscape result. A much better fit was obtained by considering the
logarithms of the same MIC values, implying that, in this case, the
interaction-free landscape is closer to a multiplicative than an
additive model (see fig.\ \ref{fig:3}). Since there is no general
theory predicting how mutational effects should combine for the
different proxies of fitness, we consistently applied the logarithmic
transformation. For all fitness measurements based on concentrations of drugs or toxins that limit growth or survival, like MIC values, the logarithms much improve the fit to the linear model. In the other cases, the logarithms did at least did not worsen the fit. Note that the MIC values for the combination of piperacillin and an inhibitor listed in \cite{Tan2011} are already the logarithms of the measurements. 

In the datasets H-J, fitness proxies are missing for several genotypes
(table \ref{table1}). The cause is either non-viability of those
genotypes (dataset H) or unobserved genotypes (dataset I-J). In the
latter case, we assume that the unobserved genotypes were missed by
chance \cite{OMaille2008}. We therefore replaced the missing
measurements by values obtained from the fitted multidimensional
linear model and subsequently performed the logarithmic
transformation. The presence of non-viable genotypes poses a problem for the log transformation of the fitness measurements. Dataset H was shown to contain non-viable \textit{A. niger} genotypes on the basis of a statistical analysis \cite{Franke2011}. A non-viable genotype would imply a logarithmic fitness equal to minus infinity. To circumvent this issue, we did not perform the log transformation for this dataset. Because the fitness values of viable genotypes (expressed in terms of relative growth rates) were fairly close to unity, taking the logarithm of the fitness values of viable genotypes would not substantially alter the results.

The results of the standardized analyses are presented in table
\ref{table2}. In short, we observe that landscapes obtained by
combining \textit{genome-wide} mutations with a known
\textit{collectively beneficial} effect are more smooth. In fact, the
landscapes A and B share these two characteristics and have the lowest
ruggedness for all four measures (see fig.\ \ref{fig:4}). The
landscapes obtained by combining mutations with a known beneficial
effect from a single gene (data sets C-F) are more rugged, while the
highest degree of ruggedness is measured in landscapes constructed
from genome-wide mutations with deleterious (data sets G and H) or
unknown (data sets I and J) effects. 
Before turning to the biological implications of these trends in
sect.\ \ref{sec:5}, we find it useful to establish the correlation
between the different measures (sect.\ \ref{subseccorrelations}) and the
fit to the RMF model (sect.\ \ref{subsecmodplusdat}). 

\begin{table}
\caption{Quantitative measures of ruggedness and epistasis for 10 empirical fitness landscapes. Except for the landscapes of size $L=4$, the reported values are averages over all possible 4-locus subgraphs. The last two lines show reference values obtained from simulations of the house-of-cards model with $L=4$ (HoC), and for a perfectly additive landscape (PA).}
\label{table2}       
\centering
\begin{threeparttable}
\centering
\begin{tabular}{llrrrrrrrrrrr}
\hline\noalign{\smallskip}
ID & Ref. & $L$ & $r/s$ & $F_1$ & $F_2$ & $F_\mathrm{sum}$ & $N_\mathrm{max}$ & $N_\mathrm{cp}$ & $f_\mathrm{r}$ & $f_\mathrm{s}$  & $f_\mathrm{mm}$ \\
\noalign{\smallskip}\hline\noalign{\smallskip}
A&\cite{Chou2011} & 4 & 0.122 & 0.989 & 0.009 & 0.011 & 1 & 24 & 0 & 0 & 1 \\
B&\cite{Khan2011} & 5 & 0.290 & 0.942 & 0.040 & 0.058 & 1.10 & 16.80 & 0.013 & 0.150& 1.92  \\
C&\cite{Lozovsky2009}\tnote{a} & 4 & 0.517 & 0.267 & 0.400 & 0.733 & 2 & 16 & 0.083 & 0.250 & 0.67 \\
D&\cite{Lozovsky2009}\tnote{b} & 4 & 0.986 & 0.537 & 0.197 & 0.463 & 2 & 10 & 0.125 & 0.458 & 0.67 \\
E&\cite{Weinreich2006} & 5 & 0.418 & 0.894 & 0.064 & 0.106 & 1.50 & 6.53 & 0.025 & 0.150 & 1.09 \\
F&\cite{Tan2011}\tnote{c} & 5 & 0.380 & 0.921 & 0.061 & 0.079 & 1.30 & 8.75 & 0.050 & 0.250 & 3.03 \\
G&\cite{Hall2010} & 6 & 1.180 & 0.658 & 0.179 & 0.342 & 2.13 & 2.10 & 0.229 & 0.358 & 3.16 \\
H&\cite{Franke2011} & 8 & 1.304 & 0.547 & 0.269 & 0.453 & 2.61 & 2.31 & 0.154 & 0.262 & 2.19 \\
I&\cite{OMaille2008}\tnote{d} & 9 & 1.317 & 0.376 & 0.368 & 0.624 & 2.66 & 2.02 & 0.240 & 0.292 & 1.71 \\
J&\cite{OMaille2008}\tnote{e} & 9 & 1.199 & 0.383 & 0.372 & 0.617 & 2.48 & 2.51 & 0.227 & 0.300 & 1.92 \\
& &  &  &  &  &  &  &  &  &  & \\
\textbullet &HoC & 4 & 2.423 & 0.267 & 0.402 & 0.733 & 3.20 & 1 & 0.333 & 0.333 & 2.20 \\
\Square &PA & 4 & 0 & 1 & 0 & 0 & 1 & 24 & 0 & 0 & 1 \\
\noalign{\smallskip}\hline
\end{tabular}
\begin{tablenotes}
\footnotesize
\item [a] Pyrimethamine resistance measurements.
\item [b] Growth rate measurements.
\item [c] Data for piperacillin resistance in the presence of a $\beta$-lactamase inhibitor; these mutations were originally selected for their beneficial effect on cefotaxime resistance \cite{Weinreich2006}.
\item [d] Relative 5-epi-Aristolochene output (main product of TEAS terpene synthase).
\item [e] Relative Premnaspirodiene output (main product of HPS terpene synthase). 
\end{tablenotes}
\end{threeparttable}
\end{table}

\subsubsection{Correlation between different measures of landscape ruggedness.}
\label{subseccorrelations}
\begin{figure}
\centerline{\includegraphics[width=\textwidth]{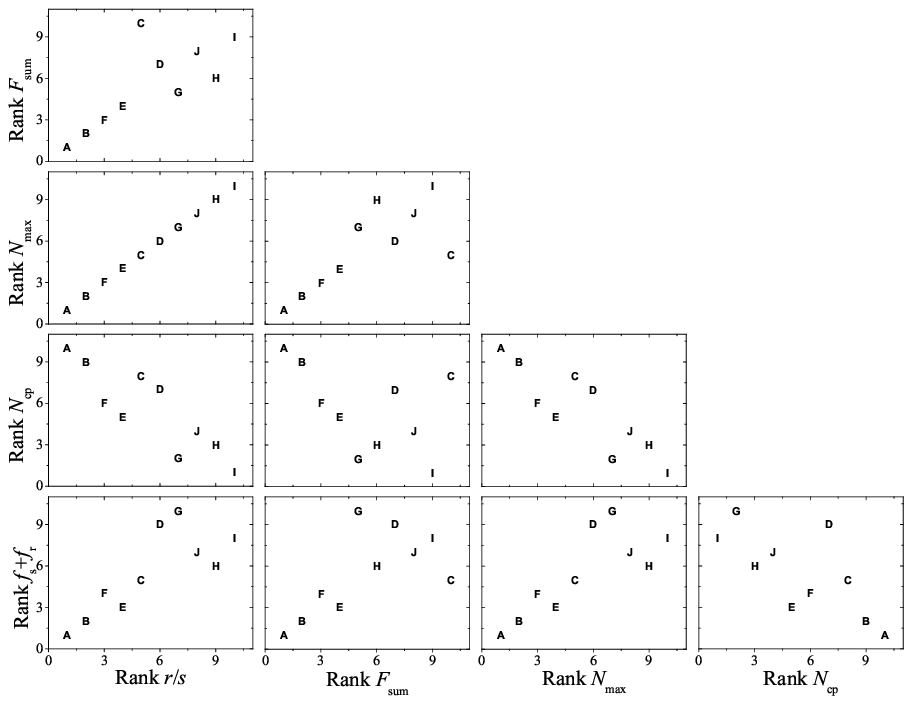}}
\caption{The ranks of the fitness landscapes specified in table \ref{table2} with respect to the studied quantifiers of epistasis are plotted against each other. In general, the quantities seem rather well correlated (see main text for a detailed discussion).}
\label{fig:4}       
\end{figure}

To investigate how well the different measures (except
$f_\mathrm{mm}$) correlate to one another, we first rank all
landscapes for each measure separately; i.e., if a landscape has the
$n$-th lowest value for a quantity, it is assigned rank $n$ with
respect to that quantity. In fig.\ \ref{fig:4}, we make pairwise plots
of these ranks for the different measures. For a perfect rank
correlation between the measures, the symbols should lie on a straight
line. In general, the different measures of ruggedness correlate well,
suggesting that these quantities all reflect the relative contribution
of epistasis in a similar way\footnote{This conclusion differs from a
  related analysis in \cite{Lobkovsky2011}, where little or no
  correlation between different roughness measures was found for a
  family of landscapes based on protein folding.}. The number of
maxima $N_\mathrm{max}$ even has a perfect rank correlation with the roughness to slope ratio $r/s$. The number of crossing paths, $N_\mathrm{cp}$, correlates somewhat less well with the other quantities. We will examine this deviation when we compare the measured values with expectations from model landscapes.

It is also instructive to compare data sets that measure different quantities using the same set of genotypes. Landscapes C and D are based on measurements of drug resistance and growth rate, respectively. The mutations in this set of genotypes were selected for their beneficial effects on resistance. Increased resistance is expected to have a trade-off in the absence of the drug and since growth rates were determined under these conditions, the included mutations are no longer beneficial. Given the tendency that landscapes from beneficial mutations are more smooth, it is not surprising that the growth rate landscape D is more rugged than the resistance landscape C. Landscapes E and F are based on a genotype with multiple mutations that is selected because of the increased resistance to a particular $\beta$-lactam antibiotic, cefotaxime. The fitness proxy used to construct landscape E is cefotaxime resistance, whereas in landscape F it is resistance to another antibiotic, piperacillin, which was not involved in the selection of the genotypes. However, the reason why they were measured in the piperacillin
environment (with inhibitor) is because resistance in this environment  showed an overall trade-off with cefotaxime resistance. Hence, the included (reverse) mutations were collectively beneficial for this environment. In contrast to landscapes C and D, the landscapes E and F turn out to be almost equally rugged.

\begin{figure}
\centerline{\includegraphics[width=\textwidth]{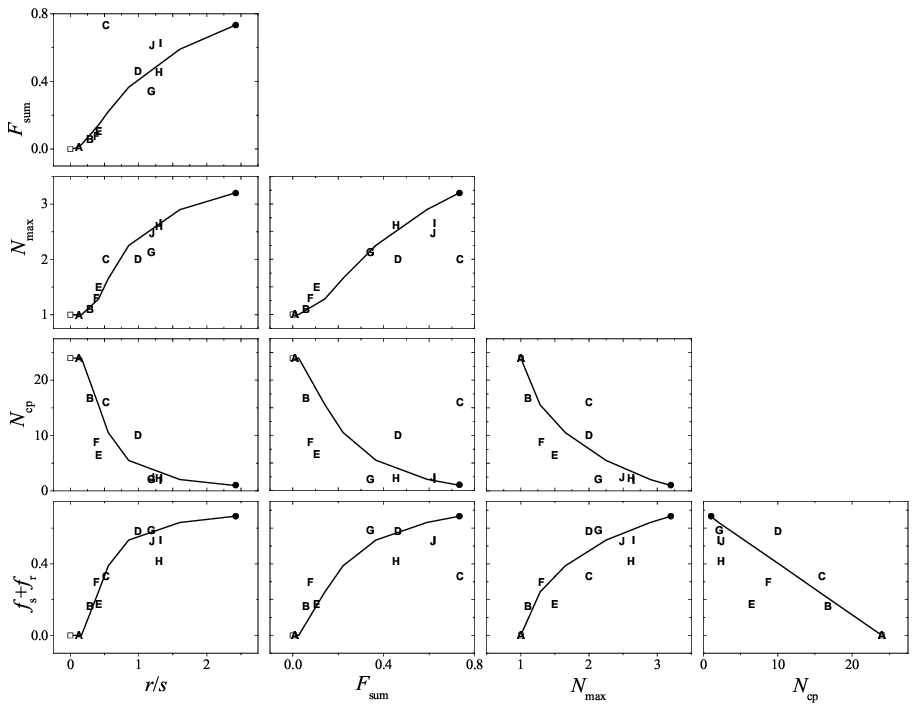}}
\caption{The measured values for the studied quantifiers of epistasis are plotted with respect to each other for the fitness landscapes specified in table \ref{table2}. The black line corresponds to values obtained numerically for RMF landscapes interpolating between a non-epistatic (open square) and a completely random (filled circle) landscape.}
\label{fig:5}       
\end{figure}

\subsubsection{Combining models and empirical data.}
\label{subsecmodplusdat}
To compare the measurements for the empirical landscapes with expectations calculated for model landscapes, we use the predictions generated by the RMF model (\ref{rmf_fitness_sum}). We thereby fixed the parameter controlling the roughness of the landscape to $a=0.01$, and calculated all measures for various choices of the slope $c$. For each pair of $a$ and $c$, the calculated values were averaged over 10,000 realizations of the 4-locus landscape. Recall that the case $c=0$ corresponds to a completely random landscape, i.e.\ to the House-of-Cards model. 
The opposite extreme of a purely additive landscape was also considered by setting $a=0$ for an arbitrary value of $c\neq 0$. 

In fig.\ \ref{fig:5}, we plot all pairwise combinations of the four epistasis measures previously included in the rank analysis of fig.\ \ref{fig:4}. The black line corresponds to the range of possible outcomes for the model landscapes, one limit corresponding to the House-of-Cards case (marked by a filled circle) and the other limit being the purely additive case (marked by an empty square). The letters represent the measurements from the experimental landscapes (see table \ref{table2}). The close correspondence between the letters and the line indicates that the RMF model captures the different ruggedness measures and their correlations observed for the experimental landscapes surprisingly well. The relatively large deviations for landscapes C and D from \cite{Lozovsky2009} are most likely due to the fact that this is a 4-locus landscape and measurements are thus based on a single observation, rather than being an average over multiple subgraphs of size $m=4$ as is the case for the larger landscapes. We also note that the number of crossing paths $N_\mathrm{cp}$ observed in the empirical landscapes appears to be systematically smaller than predicted by the RMF model, a deviation which coincides with the relatively low rank correlation of this measure compared to the other measures (fig.\ \ref{fig:4}).

\section{Discussion and outlook}
\label{sec:5}
In this review, we have first established a set of standardized measures to determine the ruggedness of fitness landscapes. We then use these measures to compare the ruggedness of ten available empirical landscapes, and to compare the empirical landscapes to predictions generated by the Rough Mount Fuji (RMF) model, a model with tunable ruggedness. Our rank analyses show that the selected measures correlate very well, and thus appear to capture the same underlying feature of the landscape. In a sense, they all capture the amount of epistasis in a particular landscape. What is quite remarkable is that not all measures are sensitive to detect all types of epistasis. The Fourier analysis and the $r/s$ ratio are sensitive to magnitude epistasis, sign epistasis and reciprocal sign epistasis. The local epistasis measure, $f_r+f_s$, and measures of accessible pathways are insensitive to magnitude epistasis, while the number of local fitness maxima is only sensitive to detect reciprocal sign epistasis. The fact that the rank correlations between the different measures are still high could either mean that the effects of sign epistasis dominate the measures that are sensitive to magnitude epistasis or that the three types of epistasis co-occur. The above also implies that the measure of epistasis among two loci ($f_r+f_s$) contains similar information as the more global measures of epistasis (involving 4 loci). Sampling local interactions can be done by detection of pairwise interactions between mutations, which is experimentally more straightforward than building multidimensional landscapes of connected genotypes. On the other hand, the Fourier analysis shows that higher-order interactions between mutations ($F_3, F_4$, etc.) play a significant role, especially in the more rugged landscapes. This information can only be detected by the construction of such landscapes.

Before we discuss which characteristics of mutations are either linked with smooth or rugged landscapes, we need to emphasize that the small number of available landscapes only allows for preliminary conclusions and that the comparison is complicated by differences in the methodologies involved to measure fitness. Nevertheless, general patterns do emerge as well as gaps in our knowledge. All empirical landscapes are relatively small in size and a specific set of mutations is used to construct the genotypes. These mutations fall into different classes. Unfortunately, few representatives are available per class, and worse: we lack any data for other classes. For example, no landscapes are available using mutations known to be beneficial by themselves in a particular wildtype, nor do we know of any studies that constructed landscapes from deleterious mutations in a single gene. This necessarily limits the interpretation and generality of our findings.

A first characteristic of a mutation which affects ruggedness is
whether the mutation is deleterious \cite{Franke2011,Hall2010} or
beneficial
\cite{Lozovsky2009,Chou2011,Khan2011,OMaille2008,Weinreich2006}. Among
the available landscapes, those that are constructed using beneficial
mutations (A-C and E-F in Table 2) are smoother than those using
deleterious mutations (G-H). Although beneficial mutations are much
rarer than deleterious ones \cite{Sanjuan2004,Perfeito2007}, most
studies focus on beneficial mutations. This seems justified given that
beneficial mutations do account for a large fraction of the mutations
that contribute to long-term evolution
\cite{Smith2002,Welch2006}. Both types of mutations are intrinsically
linked, since each fixed beneficial mutation becomes a potential
deleterious mutation when the direction of selection reverses. In that
global sense, it does not matter which type we are dealing
with. However, there is the possibility that deleterious and beneficial mutations
have intrinsically different statistical properties, including
epistatic interactions, because they sample different parts of the
local fitness landscape. Similarly, the position of the wild type is
of influence. For example, the beneficial mutations in the TEM-1
$\beta$-lactamase fitness landscape E increase resistance to the
antibiotic cefotaxime \cite{Weinreich2006}. The TEM-1 wild type incurs
a very low resistance towards this antibiotic, and the fittest
genotype has approximately a 100,000-fold higher resistance. This
clearly differs from the study by Chou et al. \cite{Chou2011}
(landscape A) in which a new metabolic pathway is introduced into a
strain of {\it Methylobacterium extorquens} and the fittest genotype
displays a $94\%$ (i.e.\ $\sim 2$-fold) fitness increase. Note also
that the empirical fitness landscapes include mutations that are known
to alter fitness (individually or collectively), and mutations with an individually neutral effect are excluded. Still, neutral mutations make up a significant portion of all available mutations \cite{EyreWalker2007}, and are known to contribute to long-term adaptation \cite{Wagner2008}.

A second characteristic that appears to influence the degree of
ruggedness is whether the individual or the combined effect
(beneficial or deleterious) is known. For example, the mutations that
were studied by Chou et al.\ \cite{Chou2011} and Khan et al.\
\cite{Khan2011} (landscape B) collectively produced a well-adapted
genotype after many generations of evolution, whereas the mutations
studied by \cite{Franke2011} (landscape H) were \textit{a priori} only known to have a deleterious effect in the wild-type background. In the first category, all intermediates are constructed between two points in genotype space that are connected by at least one accessible pathway (otherwise the higher-fitness genotype would not have been found). In the second category, the genotype that combines all mutations is not necessarily accessible from the wildtype, and does not even have to be better adapted. Consistent with an expected greater bias against (sign) epistasis in the first category, the landscapes based on genome-wide collectively beneficial mutations show less epistasis than those based on individually deleterious mutations (see table \ref{table2}). A better direct test would be a comparison between mutations with known collective or individual effect of the same fitness sign (all beneficial or all deleterious) within the same biological system, but we presently lack such data. 

A third distinction that affects ruggedness of empirical fitness landscapes is the level at which the mutations interact. The included mutations can for example affect fitness \cite{Franke2011,Khan2011}, can operate in a common genetic pathway \cite{Chou2011}, or can even be located in the same gene \cite{Lozovsky2009,Tan2011,Weinreich2006}. The landscapes constructed from beneficial mutations located in different genes (A-B) are smoother than those from beneficial mutations located in the same gene (C and E-F). When epistasis is detected between mutations in different genes, this information has traditionally been used to infer their combined contribution to a metabolic pathway. The reverse is also true: when empirical fitness landscapes combine mutations that operate in a common genetic pathway, finding epistasis becomes more likely. This becomes even more prominent when mutations are located in a single gene. Epistasis among mutations in different genes can result from functional constraints caused by interactions in a metabolic network \cite{Szathmary1993,Serge2005}, whereas intragenic epistasis can also result from structural constraints when nucleotide positions in a single gene have a combined effect on protein shape, enzyme activity, or folding-stability \cite{DePristo2005,Wang2002}. This relates to the type of epistasis that one expects to find. Magnitude epistasis is often associated with mutations in different genes in a metabolic network \cite{Sanjuan2004,Jasnos2007,Burch2004}, whereas sign epistasis is expected to occur more often between positions in a single gene \cite{Watson2011}. The expectation of a greater contribution of epistasis to landscapes based on mutations in a single gene versus in different genes is supported by our analysis (see table \ref{table2}). Note however, that compensatory mutations are often located in different genes \cite{Rokyta2011,Poon2005}, and that sign epistasis between deleterious mutations has also been shown to occur at a genome-wide scale \cite{Franke2011}.

Having in mind all complicating differences between the various fitness proxies and the diverse set of biological organisms used for testing, it is all the more surprising how well the simple model (\ref{rmf_fitness_sum}) seems to capture features of the real landscapes. However, we emphasize that we studied averaged quantities of small (i.e. 4-locus) subgraphs of the landscapes in order to standardize our measures and compare them to model predictions. Hence information contained in the full landscape might have been overlooked in these analyses. For example, the considerations in the previous paragraph suggest that the level of interaction between mutants should be distributed very inhomogeneously on large enough landscapes, consistent with predictions from metabolic models \cite{Costanzo2010,Serge2005}. This means that the landscapes can be decomposed into subgraphs, some of which contain much, others little or no epistasis. Such a decomposition would reflect the strength of interactions between specific combinations of mutations. For instance, one would expect that mutations changing the same functional part of one protein should highly influence the impact of one another on the function. On the other hand, the impact of mutations altering different proteins, which do not interact, should be independent of each other. Searching for such patterns by looking at distributions of epistasis measures instead of their mean, is a promising direction for future study.

If the existence of genetic modules with different levels of epistasis can be established empirically, it will be a challenge for future models of fitness landscapes to take this realism into account, and study its evolutionary consequences.
In fact, the $LK$ model [27, 28] was introduced with the idea of incorporating such structures. However, these models make very specific assumptions about the distribution of the size and coupling of different epistastic modules, and these should be compared to measures based on empirical landscapes. Where systematic deviations are observed, the empirical information may then be used to adapt the models. Much of the progress in understanding fitness landscapes and their evolutionary implications, therefore, depends on the availability of additional empirical landscapes from various systems -- particularly from those classes where we lack any information.

\ack
We thank Dave Hall for generously providing the original data of
landscape G. This work was supported by DFG within SFB 680
\textit{Molecular basis of evolutionary innovations}, and by
Studienstiftung des deutschen Volkes through a fellowship to JF.

\end{document}